\documentclass[10pt,aps,pra,twocolumn,superscriptaddress]{revtex4-2}

\usepackage[pages=all, color=black, position={current page.south}, placement=bottom, scale=1, opacity=1, vshift=5mm]{background}
\SetBgContents{
	\tt 
}      % copyright

\usepackage[margin=1in]{geometry} % full-width

\usepackage{amsmath}
\usepackage{amsthm}
\usepackage{amssymb}
\usepackage{times}
\usepackage{braket}
\usepackage{ulem}

\usepackage[utf8]{inputenc}
\usepackage{svg}
\usepackage{hyperref}

\usepackage[english]{babel}

\usepackage{graphicx, color}
\graphicspath{{fig/}}

\usepackage{algorithm, algpseudocode} % use algorithm and algorithmicx for typesetting algorithms
\usepackage{mathrsfs} % for \mathscr command
\usepackage{lipsum}

\begin{document}

\title{Quantum advantage for single-photon state characterization}

\author{S. N. van den Hoven}
\affiliation{MESA+ Institute for Nanotechnology, University of Twente, P.~O.~box 217, 7500 AE Enschede, The Netherlands} 

\author{M. C. Anguita}
\affiliation{MESA+ Institute for Nanotechnology, University of Twente, P.~O.~box 217, 7500 AE Enschede, The Netherlands} 

\author{S. Marzban}
\affiliation{MESA+ Institute for Nanotechnology, University of Twente, P.~O.~box 217, 7500 AE Enschede, The Netherlands} 

\author{J. J. Renema}
\affiliation{MESA+ Institute for Nanotechnology, University of Twente, P.~O.~box 217, 7500 AE Enschede, The Netherlands} 

\date{
	\today
}

\begin{abstract}
We propose a multiphoton interference protocol that characterizes the pairwise overlaps of the internal modes of single photons more efficiently than pairwise Hong-Ou-Mandel characterization experiments. We experimentally implement this protocol to characterize three photons. We show that
our implementation of the characterization protocol outperforms the pairwise Hong-Ou-Mandel characterization, even if the Hong-Ou-Mandel characterization would have been performed in a noiseless, perfect experiment. We demonstrate this via the Fisher information matrix. Surprisingly, this advantage persists at arbitrary nonzero transmission, demonstrating the viability of this protocol for real-world characterization of single photons. 

\end{abstract}

\maketitle
One of the central goals of quantum information science is to identify cases of \textit{quantum advantage}, i.e., tasks for which quantum resources provide a scaling advantage over classical ones. Paradigmatic examples include Shor's algorithm \cite{Shor_prime}, quantum computational advantage \cite{Aaronson_Arkhipov_BS}, quantum communication advantage \cite{Chakraborty_2023}, and quantum metrology below the shot-noise limit \cite{Slussarenko2017ShotNoise}. 

In linear optics, such advantages rely on the quantum interference of multiple photons arising from bosonic symmetry \cite{Fox2006QuantumOptics}. A prominent example is boson sampling, in which samples are drawn from the distribution of many single photons scattered by an interferometer \cite{Aaronson_Arkhipov_BS} and which has been experimentally realized at scale \cite{zhong2020quantum,wang2019}. Another key application is universal fault-tolerant quantum computation, where entangled resource states and projective entangling measurements are generated through multiphoton interference \cite{bartolucci2023fusion,PRXQuantum.4.020303}. In communication, these entangling operations enable the distribution of quantum information over large distances  \cite{BARRAL2025100747}.
Multiphoton interference has also enabled major advances in quantum sensing, allowing measurement sensitivities beyond the classical shot-noise limit. Milestones include Heisenberg-limited phase estimation with NOON-states \cite{PhysRevLett.75.2944, doi:10.1126/science.1138007} and squeezed-light interferometry in large-scale detectors \cite{PhysRevLett.45.75,aasi2013enhanced}. Other examples specific to linear optics include near-deterministic teleportation \cite{knill2001scheme, PhysRevLett.89.137901}, photonic fusion \cite{PhysRevLett.113.140403, PhysRevA.84.042331}, and photon distillation \cite{marshall2022distillation, sparrow2018phd_thesis, faurby2024purifying, hoch2025optimaldistillationphotonicindistinguishability,somhorst2024photon, saied2024general}.

In all of these protocols, deviations from the ideal bosonic behavior, commonly called \textit{partial distinguishability}, degrade quantum interference. This can occur because imperfections in single-photon sources impart which-way information on photons through nominally identical parts of their wavefunctions (e.g., spectrum or polarization). Partial distinguishability undermines applications, enabling classical simulation of boson sampling \cite{renema2018efficient, PhysRevA.111.052448}, or introducing mixedness during entangling operations \cite{sparrow2018phd_thesis,Saied2024AQN,PhysRevA.73.062312}.

The degree of partial distinguishability is typically determined through a characterization experiment. The standard method for pairwise characterization of two single photons is the Hong-Ou-Mandel (HOM) experiment, in which two photons enter the inputs of a balanced beamsplitter and the ratio of bunched to antibunched events is measured. This ratio directly reflects the overlap of the photons' wavefunctions \cite{hong1987measurement}. 
However, pairwise characterization of many photons quickly becomes time consuming. Moreover, small-scale quantum operations followed by classical postprocessing generally do not make optimal use of quantum resources, motivating the search for more efficient characterization protocols.

In this work, we demonstrate that characterizing the internal state of multiple single photons itself exhibits a quantum advantage. Specifically, we show that characterization based on genuine multiphoton interference outperforms conventional pairwise HOM characterization. Thus, the quantum advantage demonstrated here is that larger-scale quantum interference enables more efficient characterization than pairwise quantum interference. Given multiple continuous streams of partially distinguishable single photons and a fully tunable interferometer, our protocol characterizes all pairwise photon overlaps more efficiently than performing pairwise HOM experiments. 

Our work proceeds as follows. We first identify a protocol that outperforms HOM characterization for $n=3$ photons via numerical optimization of the Fisher information matrix (FIM). Inspecting this circuit reveals a generalization to arbitrary photon numbers, which we prove outperforms pairwise HOM characterization. We then show that this advantage persists in the presence of optical loss, a surprising result since loss typically eliminates photonic quantum advantages in computation or metrology \cite{GarciaPatron2019simulatingboson, renemaarxiv2018, PhysRevA.75.053805, PhysRevA.80.063803}. Finally, we experimentally implement the three-photon protocol and find that the FIM, inferred from the data, surpasses that obtainable in an ideal noiseless HOM characterization. This demonstrates both the practical viability of our approach and the advantage of multiphoton interference for photon characterization.

Our result differs from existing proposals to characterize partial distinguishability using multiphoton interference \cite{ji2024extensionpatternrecognitionvalidation,PhysRevX.9.011013,PRXQuantum.6.020340,Seron2024efficientvalidation,Correa_Anguita_2025,rodrigo2026benchmarking_multiphoton_interference}. These approaches either employ interferometers not optimized for information extraction or focus on global interference signatures rather than targeting each pair individually.

\begin{figure}[h!]
    \centering
    \includegraphics[width=1.0\linewidth]{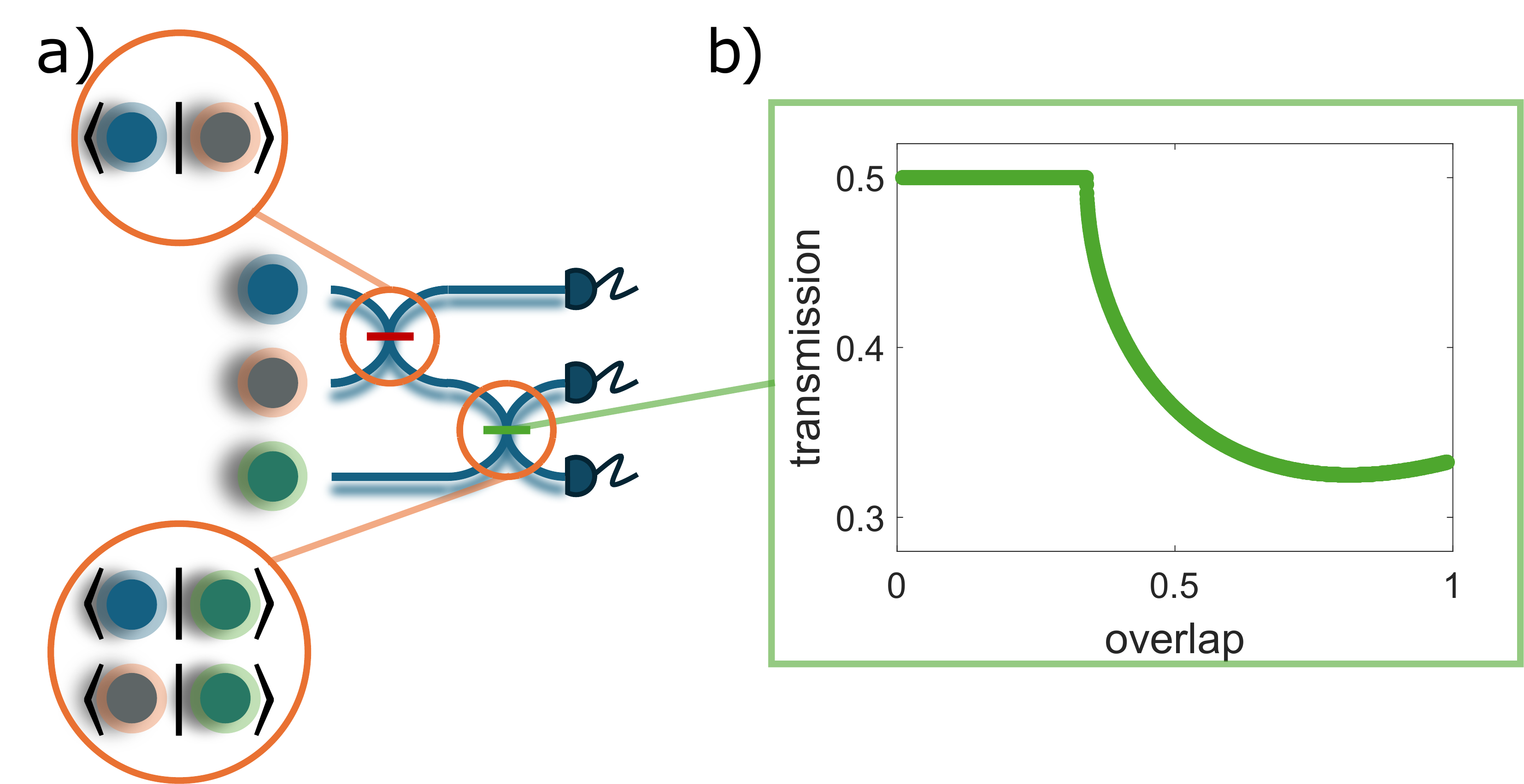}
    \caption{Improved characterization protocol. (a) Linear optical circuit consisting of a balanced beamsplitter, followed by another beamsplitter whose splitting ratio depends on the photon quality. The balanced beamsplitter effectively performs a HOM characterization on the upper two photons, while the second extracts additional information about pairwise overlaps. (b) Optimal transmission of the second beamsplitter as a function of the pairwise overlap}
    \label{fig1: better than HOM}
\end{figure}

\textit{Improved characterization for three photons}.-In a boson-sampling–type multiphoton interference experiment with partially distinguishable photons, the probability mass function (pmf) sampled by Fock-basis detection in output configuration $s$ is given by \cite{tichy2015sampling,shchesnovich2015partial}:
\begin{equation}
    P(s)=\frac{1}{\prod_i r_i! s_i!} \sum_{\sigma \in S_n} \left [ \prod^n_{j=1} \mathcal{S}_{j,\sigma(j)}\right ] \mathrm{perm}(M \circ M^{*}_{\sigma}),
    \label{eq:Tichy,multiperm}
\end{equation}
where $M$ is a submatrix of the unitary representation of the interferometer $U$, constructed by selecting rows and columns corresponding to the input and output modes ($M=U_{d(\textbf{r}),d(\textbf{s})}$). Here $d(\textbf{r})$ and $d(\textbf{s})$ represent the mode assignment lists of the input and output states, while $r_i$ and $s_i$ denote the $i^{th}$ elements of the corresponding mode occupation lists. $S_n$ denotes the symmetric group, $\mathcal{S}$ denotes the distinguishability matrix, with $\mathcal{S}_{i,j}=\langle \psi_i|\psi_j\rangle$ describing the overlap between photons $i$ and $j$ represented by their internal wave functions $\psi_i$ and $\psi_j$.  $\circ$ represents the Hadamard product, $^*$ denotes the element-wise conjugation and $M_{\sigma}$ denotes $M$ where its rows are permuted according to $\sigma$. Lastly, the permanent of matrix $M$ with shape $n \times n$ is defined as 
\begin{equation}
        \text{perm}(M) = \sum_{\sigma \in S_n}\prod_{i=1}^n M_{i, \sigma(i)}.
\end{equation}

For a given input configuration, the pmf in Eq. (\ref{eq:Tichy,multiperm}) depends on the scattering matrix $U$ and the distinguishability matrix $\mathcal{S}$. We assume that $\mathcal{S}$ contains unknown parameters to be estimated, arising from degrees of freedom unaffected by the linear optical transformation, such as temporal, spectral, or polarization modes. The matrix $U$, by contrast, is assumed to be known and under complete control.

To estimate the pmf we draw samples from the corresponding distribution and infer the parameters of interest. However, not all pmfs are equally informative. The Fisher information matrix quantifies how much information an observable random variable carries about unknown parameters, with higher values (in the matrix sense) indicating better estimation potential.
For three photons, the FIM is:
\begin{equation}
    \left[ \mathcal{I}(\theta)\right] _{ij}=\sum_{s}\left[\left(-\frac{\partial^2}{\partial\theta_i\partial\theta_j}\log{P(s)}\right) |\theta\right]P(s),
    \label{eq: FIM_definition}
\end{equation}
where $\theta=(|\langle \psi_1|\psi_2\rangle|,|\langle \psi_1|\psi_3\rangle|,|\langle \psi_2|\psi_3\rangle|)$. We do not include the triad phase \cite{PhysRevLett.118.153603Triadphase,PhysRevA.98.033805} as an unknown parameter for two reasons. First, experiments are often accurately modeled without triad phases \cite{PRXQuantum.6.020340, somhorst2023quantum}, and changing the triad phase to a nonzero value often requires considerable effort \cite{PhysRevLett.118.153603Triadphase}. Second, the HOM-experiment provides no information about the triad phase since only two photons interfere. We therefore omit it from our analysis. However, we note that another benefit of multiphoton characterization experiments is the possibility of gaining information about the triad phases. 

We exploit the additive property of the FIM, allowing the total information from multiple experiments to be obtained by summing their contributions. We choose to consider three experimental configurations, matching the number of experiments required for HOM characterization of three photons.

Each experiment is described by a distinct scattering matrix, where the $ij^{th}$ element describes the probability amplitude of the transition of a single photon from input mode $i$ to output mode $j$. For a lossless passive linear optical system acting on $N$ modes, this matrix is unitary and can be uniquely specified by $N^2$ real parameters. Common parametrizations are the Reck and Clements decompositions \cite{PhysRevLett.73.58,clements2016optimal}, where roughly half of the parameters correspond to splitting ratios of two-mode beamsplitters and the remainder to phase shifts. In boson sampling-like experiments, we perform projective measurements in the Fock-basis after the interferometer, which are insensitive to phase information. Of the $N^2$ parameters, $2N-1$ leave the pmf invariant, so only $N^2-(2N-1)=(N-1)^2$ parameters determine distinct pmfs. Here we focus on $3\times3$ matrices parametrized by three beamsplitter ratios and one phase shifter.

We used conventional numerical optimization techniques to find $12$ parameters that maximize the determinant of the sum of the FIMs from three interferometers. Maximizing the determinant corresponds to optimizing the D-optimality of our experimental design \cite{50a3d9ac-de68-352a-9395-bda5ae88bb4b}, which minimizes the uncertainty volume and is invariant under reparameterization. In general, the optimal experimental design depends on the true values of the unknown parameters. We therefore perform the optimization for different underlying values. Assuming no prior knowledge of photon quality, we optimize for the situation that the pairwise overlaps are equal. However, if information about a bias in the photon quality is available, it can be accounted for by weighting the time allocated to each experiment accordingly.

The optimization results are summarized in Fig. (\ref{fig1: better than HOM}). Subject to the constraints outlined above, we find that the optimal strategy performs three instances of the same experiment with permuted photon inputs. Two photons first interfere at a balanced beamsplitter, after which a third photon interferes with the state that comes out of one of the output arms of the balanced beamsplitter. The splitting ratio of this second beamsplitter depends on the (unknown) photon quality, as shown in Fig. (\ref{fig1: better than HOM}b). 

To better understand the behavior of Fig. (\ref{fig1: better than HOM}b), we focus on the state immediately after the balanced beamsplitter, which contains bunched and antibunched components whose amplitudes depend on the degree of partial distinguishability. For high-quality photons the bunched component dominates, yielding an optimal splitting ratio of $\frac{1}{3}$. Noteworthy, this splitting ratio, combined with a bunched component ($|021\rangle$) suppresses the $|012\rangle$ output state through destructive quantum interference. For low-quality photons, the anti-bunched component dominates, yielding an optimal splitting ratio of $\frac{1}{2}$ identical to HOM characterization.

The small non-monotonic feature near unit overlap results from the global numerical optimization of the Fisher information. While the limiting values admit simple analytical interpretations, we are not aware of a similarly simple explanation for this local feature.

\begin{figure}[t!]
    \centering
    \includegraphics[width=0.9\linewidth]{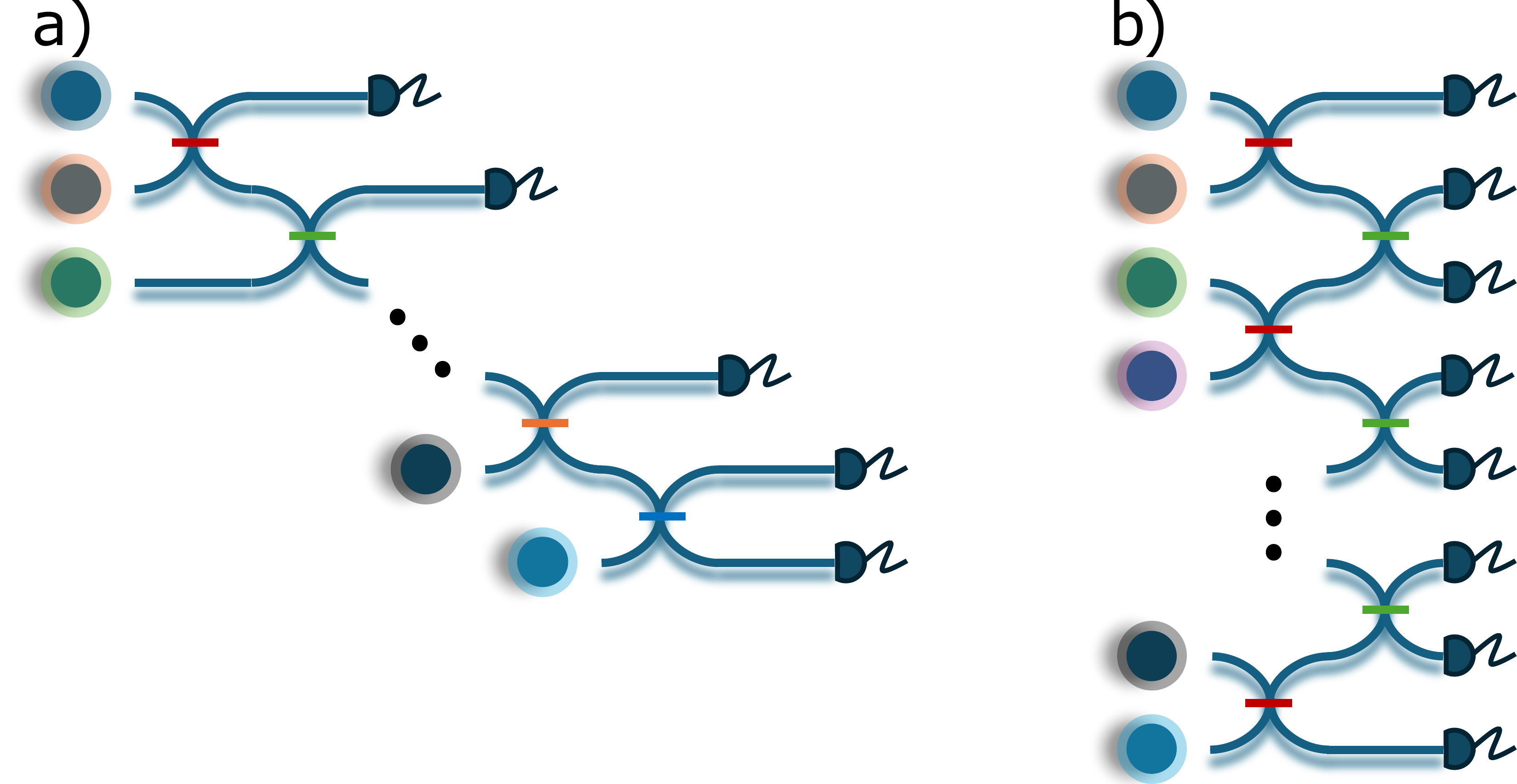}
    \caption{Generalized multiphoton characterization architectures. (a) Cascade architecture in which each stage interferes a new photon with the output of the previous one. 
    The splitting ratios can be chosen such that destructive interference suppresses specific outcomes for the dominant contribution to the output state of the previous beamsplitter
     (b) Layered architecture in which the first layer performs many HOM characterizations in parallel. The second layer introduces additional interference experiments to extract further information about pairwise overlaps. We leave the output mode of at least one balanced beamsplitter untouched (the upper one in our schematic) to determine whether bunching or antibunching events occurred at each beamsplitter in the first layer.}
    \label{fig4: generalizations}
\end{figure}

\textit{Generalization.}-We note the resemblance between the protocol in Fig. (\ref{fig1: better than HOM}) and HOM characterization. This similarity allows the scheme to be generalized to arbitrary photon numbers, outperforming HOM, although without a guarantee of optimality. The interference at the first balanced beamsplitter is fully characterized by whether the photons bunch or antibunch. Since particle-number conservation fixes the occupation of one output mode once the other is known, measuring a single output mode suffices. We can therefore interfere the state in the other output mode with a third photon, yielding additional information about pairwise overlaps. In this way, the protocol always extracts more information than a HOM characterization.
We use this observation to propose generalizations of our characterization protocol. Two examples are shown in Fig. (\ref{fig4: generalizations}). It follows from the considerations presented above that both characterization protocols
will outperform the HOM characterization for any
choice of splitting ratios (besides the beamsplitters in the first column, which must remain balanced). 

For the cascaded architecture in Fig. (\ref{fig4: generalizations}a), the number of spatial modes and beamsplitters scales linearly with the number of photons, requiring $n-1$ beamsplitters and depth $n-1$. Accessing the nonzero part of the probability mass function using threshold SNSPDs requires $2+3+\cdots+n+n=\frac{1}{2}(n^{2}+3n-2)$ detectors. This follows because successive output modes must resolve up to $2,3,\ldots,n,n$ photons, while resolving up to $k$ photons requires at least $k$ threshold detectors. The layered architecture in Fig. (\ref{fig4: generalizations}b) also uses $n-1$ beamsplitters, but the interferometric depth remains constant at 2, while the number of detectors scales linearly as $4(n-1)$.

An important structural feature of these generalized architectures is their behavior in the presence of optical loss. Unlike many quantum metrology protocols, where loss can eliminate the predicted advantage \cite{PhysRevA.75.053805, PhysRevA.80.063803}, the present scheme naturally incorporates lower-photon detection events into the characterization.

For $n=3$, for example, detection of two photons emerging from the balanced beamsplitter still yields HOM-type information about the detected pair, while three-photon events provide additional information. Thus, high-rate lower-photon data and lower-rate multiphoton events are combined within a single experiment. For heralded sources, this is straightforward, while for non-heralded sources a characterization of single-photon loss probabilities permits statistical post-processing of two-photon events. This reasoning generalizes to larger $n$, ensuring that most collected data contribute to the characterization even in lossy conditions.

In this sense, the information advantage of the protocol persists at arbitrary nonzero transmission, although loss still reduces coincidence rates. The advantage persists because lower-photon detection events remain informative, unlike in many quantum-enhanced metrology schemes.

\begin{figure*}[t]
    \centering
    \includegraphics[width=0.9\textwidth]{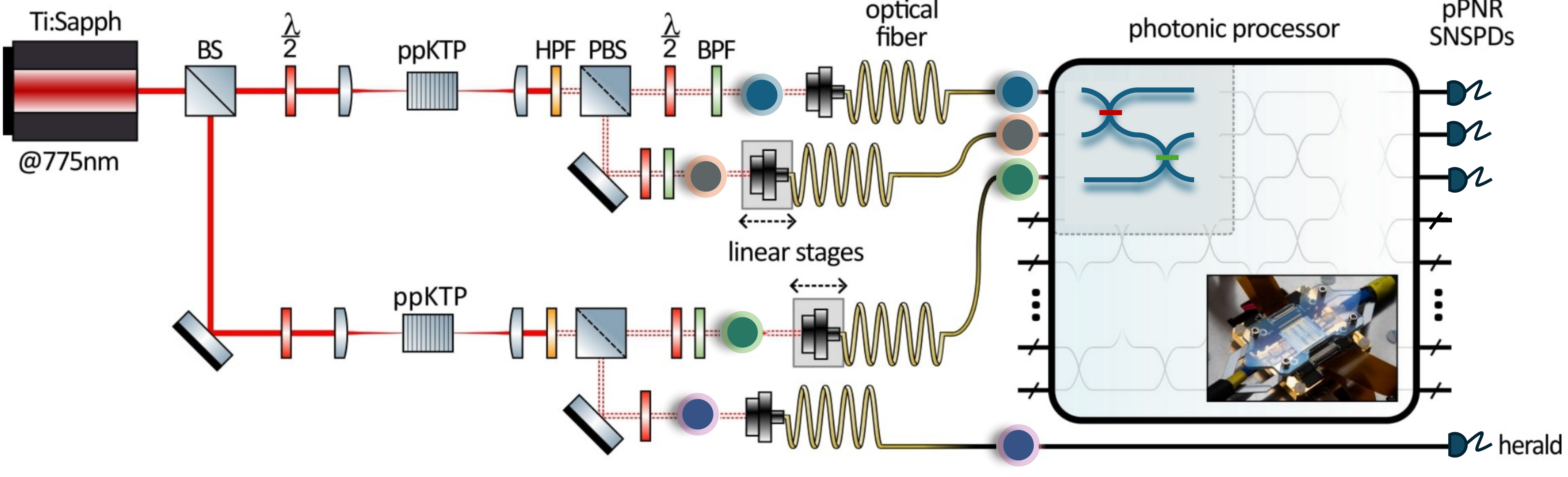}
    \caption{Experimental setup. A Ti:Sapph laser is focused onto two ppKTP crystals in parallel to produce photon pairs via type-II SPDC. After filtering with a high-pass filter (HPF), the photons are split at a polarizing beam splitter (PBS), filtered with bandpass filters (BPFs), and coupled into optical fibers. Linear stages control the path lengths and thus the relative delays of the photons, to optimize temporal overlap of the photons. The first three photons are sent to the photonic chip for the experiments, while the fourth is used as a herald. The scattering matrix presented in Fig. (\ref{fig1: better than HOM}) is programmed in the chip. A picture of the photonic chip is included as an inset. The output state is detected using a bank of quasi-photon-number-resolving (qPNR) SNSPDs.}
    \label{fig2: experimental setup}
\end{figure*}

\textit{Experiment.}-A schematic of the experimental setup, used to characterize three single photons by sampling their distribution after multiphoton interference, is shown in Fig. (\ref{fig2: experimental setup}). Single photons are generated via degenerate type-II spontaneous parametric down-conversion, featuring a pair of 2$\mathrm{mm}$ periodically poled potassium titanyl phosphate (ppKTP) crystals. The crystals are pumped by a pulsed titanium-sapphire laser ($\Delta\tau\approx$ 100 $\mathrm{fs}$), with central wavelength 775$\mathrm{nm}$ and a FWHM of $\Delta\lambda =$ 5.6$\mathrm{nm}$. The resulting two-mode squeezed vacuum (TMSV) states are sent through spectral band-pass filters of $\Delta\lambda=$ 12$\mathrm{nm}$ to enhance the purity after heralding. For one TMSV state, one mode is heralded using a superconducting nanowire single-photon detector (SNSPD), leaving a single photon (with high probability) which is sent to the integrated, tunable 12-mode interferometer. The other TMSV state is sent directly to the interferometer. We post-select on detecting three photons, projecting the TMSV onto the $|11\rangle$ state with high probability. Reducing the pump power increases the probability of projecting onto the desired $|11\rangle$ state.

The photonic processor consists of a mesh of tunable Mach-Zehnder interferometers (MZIs), integrated in low-loss silicon nitride \cite{siliconnitride_overview,taballione2021universal} in the Clements configuration \cite{clements2016optimal}. Tunability is realized via the thermo-optic effect and Joule heating. Using a carefully chosen subset of the available MZIs and feedback, we implement the $3\times3$ scattering matrix of Fig. (\ref{fig1: better than HOM}a) with amplitude fidelity ($2\sigma$ confidence interval) $F=\frac{1}{N}\textrm{Tr}(|U^{\dagger}_{target}| |U_{set}|)=1-0.000022\substack{+ 0.000016\\ -0.000074}$. We report the amplitude fidelity because input and output phases leave the measured probability distribution invariant. This metric therefore captures the physically relevant implementation accuracy. All three matrices required for our characterization differ only by permutations. Rather than programming three different matrices, we permute the input modes by manually switching fibers.

The photonic processor is followed by nine SNSPDs arranged to form three quasi-photon-number-resolving (qPNR) \cite{Feito_2009} detectors. The qPNRs are calibrated with a heralded single-photon experiment, and the translation from three-fold coincidence counts to samples of output sates is adjusted accordingly. 

\begin{figure}
    \centering
    \includegraphics[width=0.9\linewidth]{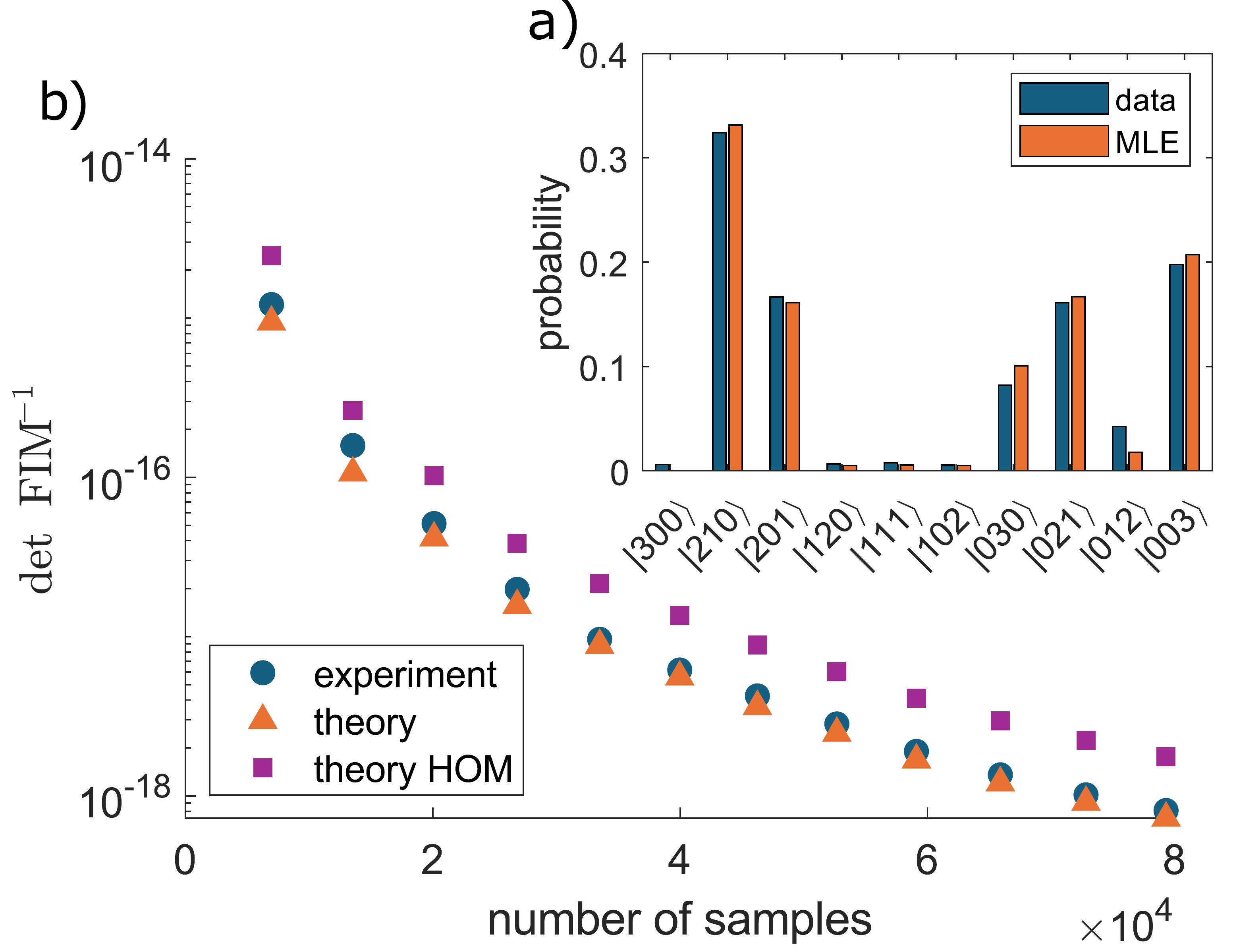}
    \caption{Experimental results. (a) The probability mass function of Fig. (\ref{fig1: better than HOM}a), showing measured data (blue) and the MLE model (orange). (b) Determinant of the inverse Fisher information matrix versus number of samples. Experimental results (blue circles) are compared with ideal HOM characterization (purple squares) and an ideal implementation of our protocol (orange triangles), demonstrating a clear information advantage.}
    \label{fig3: experimental results}
\end{figure}

The measured samples from the three different experimental configurations are used in a maximum likelihood estimate (MLE) of the parameter set $\theta = (|\langle \psi_1|\psi_2\rangle|,|\langle \psi_1|\psi_3\rangle|, |\langle \psi_2|\psi_3\rangle|, \phi_{\mathrm{triad}}, t_1,t_2,t_3,\alpha)$. Here, $t_1$, $t_2$ and $t_3$ represent the beamsplitters' splitting ratios in a 3-mode Clements (or Reck) decomposition, $\alpha$ describes the phase shifter. Although we perform three separate experiments, we assume that the samples are drawn from pmfs that are parametrized by the same set. We used the same interferometer for all experiments, and the interferometer is stable over time. Fig. (\ref{fig3: experimental results}a) shows the measured (and normalized) counts together with the pmf resulting from the MLE for one of the three experiments. To avoid cherry picking, we show the pmf with the largest total variational distance (tvd) between model and experiment. The featured pmf has a tvd of $0.041$, the other two experiments yield tvds of $0.024$ and $0.020$, see Supplementary Material \cite{SM1}. 

Fig. (\ref{fig3: experimental results}b) compares the determinants of the inverses of three FIMs as a function of the number of chronologically acquired samples. The first FIM is computed from our experimental data using Eq. (\ref{eq: FIM_definition}). The second represents an ideal HOM characterization (noiseless and perfectly balanced). The third represents an ideal implementation of our protocol, which is included to illustrate the effect of experimental noise. The results show that, for the same number of samples, our protocol provides more information about pairwise overlaps than an ideal HOM-based approach. The observed FIM is also close to the ideal value, demonstrating robustness to experimental noise. Numerical simulations of interferometric noise (Supplemental Material \cite{SM2}) confirm that the information advantage persists under realistic fluctuations of the beamsplitter transmissivities. For the photon quality realized in our experiment, ideal HOM characterizations would require $1.30$ times more samples to reach the same uncertainty volume (in the D-optimal sense) as our implemented protocol.

\textit{Discussion and conclusion.}-We have introduced a protocol for characterizing pairwise overlaps of single photons. This protocol outperforms the conventional and broadly used HOM characterization, quantified in terms of the Fisher information matrix. It can be understood as a natural multiphoton extension of HOM characterization and applies to arbitrary photon numbers, a regime that is becoming increasingly accessible in modern multiphoton experiments. For the case of three photons, we show that our protocol is optimal. This demonstrates that multiphoton interference contains information about partial distinguishability that cannot be reconstructed from any collection of pairwise overlap measurements. 

More generally, this reflects a broader shift in quantum science toward regimes where many-body systems become experimentally accessible, potentially requiring characterization strategies beyond pairwise experiments.

Importantly, we find that this advantage persists at arbitrary nonzero transmission, demonstrating the viability of this protocol for real-world characterization of single photons. In contrast, losses in metrological protocols often destroy the practical advantage predicted theoretically \cite{PhysRevA.75.053805, PhysRevA.80.063803}.

We experimentally implement this protocol for three photons and observe an information matrix exceeding that obtainable from a sequence of ideal, noiseless HOM characterizations, confirming the operational relevance of the predicted advantage.

The protocol requires a somewhat more complex interferometer and additional detectors than standard HOM characterization. However, the added interferometric complexity is modest, extending a HOM setup with additional beamsplitters suffices. Limited detector availability may nevertheless motivate the use of HOM characterization, which can be realized with as few as four detectors using qPNRDs, or even two at the cost of doubling the required samples.

\textit{Acknowledgments.}-We thank F.H.B. Somhorst, N. Walk and C. Toebes for scientific discussions. This research is supported by the PhotonDelta National Growth Fund program. This publication is part of the project At the Quantum Edge (VI. Vidi.223.075) of the research programme VIDI which is financed by the Dutch Research Council (NWO).

\textit{Data availability.}-All experimental and simulated data used in this study are available in
the 4TU.ResearchData database. \cite{doi.org/10.4121/aea16cd4-1a57-442d-b06c-643fcdab6393.v1}

\bibliographystyle{apsrev}

\bibliography{refs.bib}

\end{document}